\documentstyle[12pt]{article}
\begin{document}
\baselineskip=24pt
\begin{center}
{\large\bf A note on magnetic-field induced level-density condensation in a
two-dimensional electron gas with point scatterers}
\vspace{.30cm}

{\bf N. Kumar and Nivedita Deo}\\
{\em Raman Research Institute, C V Raman Avenue, Bangalore 560080, India}
\vspace{0.50cm}
\end{center}

\noindent
{\bf Abstract}

The density-of-states (DOS) for a magnetized (B) two-dimensional electron
gas (2DEG) containing point scatterers of arbitrary strengths,
concentration ($n_s$) and distribution is analyzed. It is shown from
the first principles that for  \(n_s \leq B/\Phi_o \equiv n_B\), the areal
density of flux quanta \(\Phi_o \equiv hc/e\), the DOS retains the extensive degeneracy
characteristic of the Landau levels, but reduced by a factor \((1 -
{n_s}/{n_B})\). This elementary but exact result gives a level
condensation for  magnetic field \(B > n_s\Phi_o\), as first noted by Br\'{e}zin
{\em et al.}. Its implications for the Integral Quantum Hall Effect and
for Random Matrix Theory are pointed out.



\baselineskip=12pt
\begin{tabbing}
{\it Corresponding author}: \=Prof. N.Kumar\\
\>Raman Research Institute\\
\>Bangalore 560080, India\\
\>Telephone: (+91 80) 331 1012\\
\>Fax:~~~~~~~~~~(+91 80) 334 0492\\
\>e-mail: nkumar@rri.ernet.in
\end{tabbing}

\baselineskip=24pt

\newpage
A two-dimensional electron system in an external magnetic field (B)
applied perpendicular to the plane, and with randomly distributed and
strengthed impurities (disorder) is one of the most extensively studied
low-dimensional systems, known best for the robust  phenomenon of Integral
Quantum Hall Effect (IQHE) [1].  The simplest model  here consists of a
two-dimensional electron gas (2DEG), i.e., without electron-electron
interaction, spin-polarized in a strong uniform perpendicular magnetic field with
uncorrelated point-like scatterers having a Poissonian distribution in the
plane. It is described by the one-electron Hamiltonian [2]
\begin{equation}
H = \frac{1}{2m} \left( -i\hbar \nabla  - \frac{e}{c} A\right)^2 \,\,+\,\,
\sum_{i=1}^{N_s} V_i \delta(r - R_i) \equiv H_o + V
\end{equation}
\noindent
with the vector potential {\bf A} = B(0, -x) in the Landau gauge. In the
following we consider the general case of arbitrary spatial positions   
(\{$R_i$\}) and strengths (\{$V_i$\}) of the point scatterers. The mean
concentration  of scatterers
\(n_s = N_s/A\), with the area \(A = L_x . L_y\).  Thermodynamic
limit \(L_x, L_y \rightarrow \infty  \) will be assumed throughout.

The Hamiltonian $H_o$ describes the well known Landau problem having
eigenfunctions \(\Phi_{nk}(\bf r)\) and the  corresponding eigenvalues \(E_{n} =
(n + \frac{1}{2}) \hbar w_c\),  labelled by the Landau-level index  \(n
= 0, 1, 2, ... \),  and the quasi-continuous  momentum label $k$ taking
on \(N_B(=eB/hc)A\) values. Here \(w_c = eB/mc\), the cyclotron frequency.
Each Landau level is extensively degenerate with an areal density \(n_B =
B/\Phi_o\), where $\Phi_o = hc/e$ is the magnetic flux quantum.

The unperturbed DOS is thus given by
\begin{equation}
N_o(E;B) = (B/\Phi_o) \sum_{n=0}^{\infty} \delta\big(E - \hbar
w_c(n+\frac{1}{2})\big)
\end{equation}
We will now show that the exact DOS $N(E;B)$  for the disordered system 
continues to have an extensive degeneracy for $B > B_c = n_s\Phi_o$,
characteristic of the Landau system, but reduced by a factor \((1 -
n_s/n_B)\). This implies level condensation as the field exceeds the
critical field $B_c$. The uncondensed fraction $n_s/n_B$ forms a diffuse
background to which the Wigner semi-circular law of Random Matrix Theory
(RMT) [3] should  apply.

With \(N_B > N_s\),  consider the n$^{th}$ Landau level of $H_o$ and
construct the $N_B$ degenerate linear combinations 
\begin{equation}
\psi_n^m({\bf r}) = \sum_{k=1}^{N_B} C_{nk}^m\,\,\phi_{nk} ({\bf r}),\,\,\,\,\,\,\,m=1,2,...,N_B
\end{equation}
as the new bases. Now, choose the coefficients
\(\{C_{nk}^m\}\) such that \(\psi_n^m(\bf r) \) = 0 for \(r = R_i\), i.e.,
\(\psi_{n}^m(\bf r)\) has nodes at the sites of the point scatterers,
\begin{equation}
\psi_n^m (r = R_i) \equiv \sum_{k=1}^{N_B} C_{nk}^m \phi_{nk}(r = R_i)
= 0
\end{equation}
for $m = 1,2,...., N_s$.

Inasmuch as there are $N_s$ conditions (equations) and $N_B$ unknown
coefficients, we are left with a sub-space of dimension ($N_B - N_s$)
spanned by the linear combinations $\psi _n^m$ that vanish at \{$R_i$\}.
Further, 
since these linear combinations vanish at the impurity sites \{$R_i$\},
they remain unaffected by the point scatterers [4].  Thus, there will be
degenerate ($N_B -
N_s$) eigenstates (per Landau level) unaffected by the disorder.
Accordingly, we have the perturbed DOS
\begin{equation}
N(E;B) = \sum_{n} \left(1 - \frac{n_s}{n_B}\right) \delta(E - E_n) +
N_{diffuse}(E)\,\,. 
\end{equation}
Thus, there is a condensation of level-density for a strong magnetic field. The
condensation vanishes for \(B \leq B_c = n_s\Phi_o\). The uncondensed
fraction distributed as a band \(N_{diffuse}(E)\) corresponds to the states
affected by disorder. For a clean sample with \(n_s \approx 10^{11}\,\,{\rm
cm}^{-2}\), we have the critical field \(B_c \approx 20 \)kG. The
condensed eigenstates (extended) at the centre(s) of the Landau band(s)
are believed to be important for IQHE [5].

It is now apt to compare the
DOS in Eq.(5) to the one obtained perturbatively by  Br\'{e}zin {et al.} [2]
(their Eq.50a) using the supersymmetric formalism.  The condensate part
matches exactly. However, it must be noted that their result was obtained
specifically for the Poissonian distribution after averaging over the
scattering impurity strengths, and for the single lowest Landau level
(LLL). The present 
first-principle result for the condensation holds for the most general
distribution of point scatters, and, in fact, without restriction to LLL.
Physically, point 
scatterers would imply the range of scattering potential much smaller than
the magnetic length \(\big(= \sqrt{\hbar c/eB}\big)\). For the diffuse part of
DOS, however, we have no 
exact results here. But, the following argument should hold in the
strong-field limit $-$ if the inter-Landau level scattering is neglected
as in ref. 3. (Though this may not be a very good approximation for the case of
{\em abrupt}, short-ranged scatterers [6].)
As the uncondensed fraction $n_s/n_B$ of the total states in the 
 LLL is subject to random scattering, one may expect the
results of the RMT to hold, and hence, in particular, we expect [3]
\begin{equation}
\frac{N_{diffuse}^{LLL}}{N_B} (E)  =  \frac{n_s}{2\pi\Delta n_B} \sqrt{4 - \left(\frac{E-E_o}{\Delta}\right)^2}
\end{equation}
\noindent
We would like to
emphasize, however, that while the condensate part of the density of
states Eq.(5) is exact for the point scatterer model, the
diffuse part \(N_{diffuse}\) is some non-universal background. Only for
Gaussian RMT it assumes the semicircular form. 
We note here that a DOS with a delta-function and a semi-circular
background has been obtained earlier [9] using RMT.
In point of fact, in the absence of inter-Landau level scattering, all
eigenstates can be readily shown to be extended  [6,7] for the
quantum-Hall-effect systems. And the extended states are, of course, a
necessary condition for the RMT to hold. We can therefore treat Eqs.(5)
\& (6) as providing the input density of states for constructing the
effective potential for the RMT unitary ensemble with the parameter
($n_s/n_B$) tunable by varying the magnetic field. We can then invoke the
known expressions for the various correlation functions evaluated by Nagao
{\em et al.} [10].  This then provides us with the possibility of
studying these correlation functions as function of magnetic  field for
the quantum Hall regime.  In particular, the coefficient `a' (strength of
the log potential in their Eqs. (2.2) \& (2.3)) must vanish and stay zero
for $B < B_c$. To our knowledge, this is the first time that a condensed
matter system has been identified for the RMT-model potential giving the
semi-circle plus delta function DOS. The non-condensed diffuse DOS
obtained by  Br\'{e}zin{\em et al.} is, however, qualitatively different from
the above semi-circular law. 

We would like to conclude with a remark of direct relevance to the Hall
conductance plateaus in the context of the above model (i.e., within the lowest
Landau Level subspace). Normally, a Hall plateau represents the gap
between the two mobility edges separating the adjacent Landau levels. As
the disorder increases, the mobility edges are expected to move inward towards the
respective Landau level centres, and hence the width of the plateau should
increase, making the Hall steps steeper. In the present model, however,
the opposite should happen.  As all the states are extended, the plateaus
can arise only due to the DOS gaps between the adjacent Landau levels [7].
Therefore, increasing the disorder strength would decrease the
plateau-width as the DOS gap decreases because of the increasing width of
the diffuse band ($N_{diffuse}(E)$) about the Landau levels. To the extent 
that the model of  Br\'{e}zin{\em et al.} is a good approximation in the
high-field limit, the above opposite behaviour is inescapable.\\

\noindent
{\bf Acknowledgements}

N.D. would like to thank E. Br\'{e}zin, R.N.Bhatt, F. David and S. Jain
for discussions.

\end{document}